\documentstyle[12pt]{article}

\newcommand{\Ima}{\hbox{Im}}

\newcommand{\Od}{{\cal O}}

\begin{document}
\input feynman
\input epsf

\thispagestyle{empty}
\hfill    LBNL-39090

\hfill    UCM-FT 5/96

\hfill    July 1996

\begin{center}
{\LARGE \bf  Higgs physics in the Large $N$ Limit}
\footnote{This work was supported by the Director, Office of Energy
Research, Office of High Energy and Nuclear Physics, Division of High
Energy Physics of the U.S. Department of Energy under Contract
DE-AC03-76SF00098.}
\\

\vskip 1cm

{\bf A. Dobado}\footnotemark[2]\\
\small
Departamento de F\'{\i}sica Te\'orica.\\
Universidad Complutense de Madrid.
28040 Madrid, Spain \\

\vskip .5 cm

\normalsize
{\bf J. Morales}\\
\small
Departamento de F\'{\i}sica.
Universidad Nacional de Colombia\\
A.A. 5997 Bogot\'a, Colombia

\vskip .5 cm

\normalsize
{\bf J.R. Pel\'aez}\footnotemark[3]\\
\small
Theoretical Physics Group. Lawrence Berkeley Laboratory\\
University of California.
Berkeley, California 94720. USA.\\

\vskip .5 cm

\normalsize
{\bf M.T. Urdiales}\\
\small
Departamento de F\'{\i}sica Te\'orica.\\
Universidad Aut\'onoma de Madrid.
 28049 Madrid. Spain\\
\end{center}

\vskip .8 cm

\begin{abstract}
In this paper we study the large $N$ limit of the Standard Model  Higgs
sector with  $N\lambda
$, $Ng^2$ and $Ng'^2$ constant and $N$ being the number of would-be Goldstone bosons. 
Despite the simplicity of this method at leading order, 
its results satisfy
simultaneously important requirements
such as unitarity and the low-energy theorems in contrast with other more 
conventional approaches. Moreover, it is fully compatible with the Equivalence
Theorem and it yields a consistent description of the Higgs boson 
mass and width. Finally we have
also included a phenomenological discussion concerning the applications of
this method to the LHC.\\

\vskip .3cm

\end{abstract}

\footnotetext[2]{E-mail:dobado@eucmax.sim.ucm.es}
\footnotetext[3]{
Complutense del Amo fellow. On leave of absence from:
Departamento de F\'{\i}sica Te\'orica. Universidad Complutense.
28040 Madrid, Spain. E-mail: pelaez@theor3.lbl.gov, pelaez@vxcern.cern.ch}
\newpage

\newpage

%
%
%

\section{Introduction}

As it is well known, the most popular theoretical description of the
Symmetry Breaking Sector (SBS) of the Standard Model (SM), 
is given by the Minimal Standard Model (MSM) which is nothing but an
$SU(2)_L\times U(1)_Y$ gauged linear sigma model. Indeed, the hidden 
sector displays an $SU(2)_L\times SU(2)_R$ {\em global} symmetry 
which is spontaneously broken down
to $SU(2)_{L+R}$. This mechanism is responsible for the
 the spontaneous breaking of the 
{\em gauge} symmetries of the complete model. 
In this scheme we have three would-be Goldstone bosons, which
will give masses to the  $W^{+}, W^{-}$ and $Z^0$ through the Higgs mechanism.
They parametrize the space spanned by the three broken generators, i.e. the 
coset \begin{equation}
\frac{SU(2)_L\times SU(2)_R}{SU(2)_{L+R}}\simeq \frac{O(4)}{O(3)}
\end{equation}
There is however a particle
which survives the Higgs mechanism, which is known as 
the Higgs boson. This particle is the only missing piece of the MSM
and for this reason it is very important to be able to predict its behaviour
in order to confirm or reject the MSM experimentally.

At tree level
the dynamics of the Higgs sector is controlled by its self-coupling
$\lambda$. In fact, its mass is related with this constant by the simple
equation $M^2=2\lambda v^2$, where $v \simeq 250$ GeV is the vacuum
expectation value. Notice that this equation suggests that a heavy Higgs will give
rise to a strongly interacting Higgs sector (see \cite{Cha} for review). However,
it should be kept in mind that for large $\lambda$ the above equation does
not hold any more, since perturbation theory is not reliable. As a matter
of fact, the tree
level amplitudes break unitarity for Higgs masses around $1$ TeV \cite{LQT}.

Therefore, it seems clear that a more complex dynamics should 
emerge for large coupling.
At the same time, there are strong hints supporting
the triviality of the minimal Higgs sector (see \cite{Ca} for a review),
which means that it 
should be considered as some kind of effective theory which can be
applied only for energies well below some cutoff $\Lambda$. In such case, the
Higgs mass becomes a decreasing function of this cutoff in such a way that, at
some point around 1 TeV, one has $M \simeq \Lambda$. This fact is usually
interpreted as an upper bound for the  Higgs mass,
 since it should not be larger than the cutoff $\Lambda$
of the effective theory.

From
the practical point of view the natural place to probe this  dynamics is
gauge boson scattering. As it is well known, the longitudinal components of the
$W^{+}, W^{-}$ and $Z^0$ gauge bosons are related with the three would-be
Goldstone bosons. The precise relation is given by the Equivalence Theorem (ET)
\cite{ET,LQT}, which states that at high energies the 
$S$-matrix elements of longitudinal gauge bosons
are the same as those of their corresponding GB.
This theorem is very useful since it is far easier to
work with the would-be Goldstone bosons than with gauge bosons. The
ET has been widely used in many studies concerning the
discovery of the Higgs boson at the future Large Hadron
Collider (LHC) (see \cite{LHC} and references therein). With its help and
at {\em lowest order} in the $g$ and $g'$ $SU(2)_L\times U(1)_Y$ gauge couplings,
it is possible to reduce the study of longitudinally polarized gauge boson
dynamics to the {\em non-gauged} $O(4)/O(3)$ linear sigma  model.

Nevertheless, the tree
level, or even the one-loop approximation \cite{1L}, does not provide a complete
description of the expected behaviour of the physical Higgs \cite{HH}. This
is due to the fact that, in the strong interacting regime, i.e. for large
$\lambda$, the standard perturbation theory does not work. In particular it
is not able to reproduce properly the position and the width of a heavy
Higgs. For this reason some non-perturbative techniques have been studied in the
literature like the $N/D$ method  (see \cite{LQT} and \cite{N/D}) or the Pad\'e
approximants \cite{Pade}. 

An alternative approach to those listed above 
is the so called large $N$ limit
\cite{LN}. The main idea is to extend the $O(4)/O(3)$ 
symmetry breaking pattern of the
linear sigma model to $O(N+1)/O(N)$. Once this is done, 
the amplitudes are obtained to lowest order in the $1/N$ parameter
\cite{LargeN}. The relevant point is that in this simple manner 
it is possible to study some
properties of the Higgs dynamics, which are expected theoretically, but
that cannot
be reproduced with more conventional techniques. In particular, the
would-be Goldstone boson elastic scattering amplitudes are unitary (up to
$\Od(1/N^2)$ corrections) and satisfy the Weinberg low-energy theorems coming from
the $O(N)$ symmetry \cite{LET}. Moreover, the Higgs propagator has a pole in
the second Riemann sheet that has to be understood as  the physical Higgs. The
position of this pole is a function of the renormalized Higgs mass $M$ but its
real  part is never bigger than some value around 1.5 TeV, even in the  $M\rightarrow\infty$
limit.
The fact that there is a saturation value for the Higgs mass
is consistent with the assumed triviality of the $O(4)/O(3)$ model and has
also been found in other non-perturbative approaches like the above mentioned
$N/D$ method  or the Pad\'e approximants.

In this work we have applied the large $N$ techniques
to an $O(N+1)/O(N)$ linear sigma model which has been gauged
with the $SU(2)_L \times U(1)_Y$ symmetry of the SM. 
 The aim of this generalization is
twofold. First it will be possible to compute the elastic gauge boson
scattering amplitudes {\em without} using the ET. This is very
important since then we can apply our results at low energies too. 
Nevertheless we show how the ET works remarkably well in the
large $N$ approach, which is also a nice check 
of our computations at high energies. Second, by gauging the linear 
sigma model, we are able to
to include systematically the $g$ and $g'$
corrections keeping at the same time the very good
properties of the standard large $N$ limit. We will show that this approach
is very easy to implement and for this reason it is
appropriate to describe the Higgs phenomenology at the LHC.

 The
plan of the paper goes as follows. In section two we introduce the $SU(2)_L
\times U(1)_Y$ gauged $O(N+1)/O(N)$ linear sigma model.  In section three we
study the main properties of the physical  Higgs boson in this approximation.
In the fourth we check our method with
the equivalence theorem and how it is satisfied in  the
large $N$ limit. In the fifth we show our numerical results, which are relevant
for the LHC phenomenology. Finally in section six we give the main
conclusions of our work.

\section{The large $N$ limit of the Higgs sector}

 We start from the  $SU(2)_L \times U(1)_Y$
gauged version of the linear sigma model $SU(2)_L \times SU(2)_R/SU(2)_{L+R}
\simeq O(4)/O(3)$
generalized to the coset $O(N+1)/O(N)$. The
classical lagrangian is then given by
\begin{equation}
{\cal L}={\cal L}_{YM}+\frac{1}{2}(D_{\mu}\Phi)^T
D ^{\mu}\Phi-V(\Phi^2) ,
\label{lagra}
\end{equation}
with $\Phi^T=(\pi^1,\pi^2,...,\pi^N,\sigma)$
and $\Phi ^2=\Phi^T\Phi$. As usual, ${\cal L}_{YM}$ is the standard $SU(2)_L  \times
U(1)_Y$ Yang-Mills term
and the  covariant derivatives are defined as
\begin{equation}
D _{\mu}\Phi=\partial _{\mu}\Phi-igT_a^LW_{\mu}^a
\Phi+ig'T^YB_{\mu}\Phi,
\end{equation}
where the $SU(2)_L$ and the $U(1)_Y$ generators are
$T_a^L=-(i/2)M_a^L$ and $T^Y=-(i/2)M^Y$ with
\begin{eqnarray*}
M_1^L=
\left(
\begin{array}{ccccc}
 0 & 0 & 0 & ...& - \\
 0 & 0 & - & ...& 0 \\
  0 & + & 0 & ...& 0 \\
   ... &  &  & &  \\
    + & 0 & 0 & ...& 0 \\
\end{array}    \right)  &,&    
    M_2^L=
\left(\begin{array}{ccccc}
 0 & 0 & + & ...& 0 \\
 0 & 0 & 0 & ...& - \\
  - & 0 & 0 & ...& 0 \\
   ... &  &  & &  \\
    0 & + & 0 & ...& 0 \\
    \end{array}    \right)     \\   \nonumber
    M_3^L=
\left(\begin{array}{ccccc}
 0 & + & 0 & ...& 0 \\
 - & 0 & 0 & ...& 0 \\
  0 & 0 & 0 & ...& + \\
   ... &  &  & &  \\
    0 & 0 & - & ...& 0 \\
\end{array} \right) &,&
M^Y=
\left(\begin{array}{ccccc}
 0 & + & 0 & ...& 0 \\
 - & 0 & 0 & ...& 0 \\
  0 & 0 & 0 & ...& - \\
   ... &  &  & &  \\
    0 & 0 & + & ...& 0 \\
\end{array} \right).
\end{eqnarray*}
where all the non written entries vanish.
 The potential is given by
 \begin{equation}
 V(\Phi^2)=-\mu^2\Phi^2+\frac{\lambda}{4}(\Phi^2)^2,
 \end{equation}
whose tree level minimum is reached whenever $\Phi^2
=v^2=NF^2=2\mu^2/\lambda$. As a consequence once we choose a vacuum
to quantize the theory,
the original $O(N+1)$ symmetry will be broken down to $O(N)$.
With the standard choice $\Phi^T_{vac}=(0,0,...,0,v)$ 
and defining the
Higgs field as $H=\sigma-v$, we can write
 \begin{equation}
V(\pi,H)=-\frac{1}{2}M_H H^2-\frac{\lambda}{4}(\pi^2+H^2)^2-\lambda v
H(\pi^2+H^2) , 
 \end{equation}
 where the tree level Higgs mass is given by $M^2_H=2\lambda v^2$.

In order to obtain a well defined perturbation theory, one  has
to add a gauge fixing and a Faddeev-Popov term
to the lagrangian in Eq.\ref{lagra}. 
As far as we are dealing with a gauge theory which is spontaneously
broken, it is specially useful to choose an $R_{\xi}$ gauge, where now 
$\pi^1,\pi^2$ and $\pi^3$ can be directly identified with
the would-be Goldstone bosons. With the complete lagrangian at hand it
is possible to derive the Feynman rules
following the  usual procedures. For convenience, we will be working
all the time in the Landau gauge, which simplifies the calculations
since the ghosts do not couple directly to the  $\pi^a$ fields and 
their propagator does not have a mass term.

\section{The Higgs mass and width}

In order to study the main properties of the Higgs resonance in the large $N$
limit of the model defined above, we will start by setting
$g=g'=0$, i.e. we will turn off the gauge interactions. Thus
the only fields we have to consider are the $N$ Goldstone bosons
$\pi^a$ and the Higgs $H$. 
Thanks to the remaining $O(N)$ symmetry as well as to crossing symmetry,
the  scattering amplitude for the process
$\pi^a\pi^b\rightarrow\pi^c\pi^d$ can be written as
\begin{equation}
T_{abcd}(s,t,u)=A(s,t,u)\delta_{ab}\delta_{cd}+A(t,s,u)\delta_{ac}\delta_{bd}
+A(u,t,s)\delta_{ad}\delta_{bc}. 
\end{equation}

\begin{figure}

\begin{picture}(20000,20000)

\THICKLINES
\bigphotons

\drawline\fermion[\SE\REG](7500,20000)[6000]
\global\advance\pmidx by 150
\put (\pmidx,\pmidy){\circle*{700}}
\global\advance\pmidx by -150
\drawline\fermion[\NE\REG](\pmidx,\pmidy)[3000]
\drawline\fermion[\SW\REG](\pfrontx,\pfronty)[3000]

\put (4000,\pfronty){\small a)}

\put (6700,20000){$\pi$}
\put (6700,15000){$\pi$}
\put (12300,20000){$\pi$}
\put (12300,15000){$\pi$}

\global\advance\pfrontx by 5000
\global\advance\pfronty by -500

\put (\pfrontx,\pfronty){$\equiv$}
\global\advance\pfronty by 500

\global\advance\pfrontx by 5000

\drawline\fermion[\NE\REG](\pfrontx,\pfronty)[3000]
\drawline\fermion[\SE\REG](\pfrontx,\pfronty)[3000]
\drawline\fermion[\SW\REG](\pfrontx,\pfronty)[3000]
\drawline\fermion[\NW\REG](\pfrontx,\pfronty)[3000]

\put (16700,20000){$\pi$}
\put (16700,15000){$\pi$}
\put (22300,20000){$\pi$}
\put (22300,15000){$\pi$}
 
\global\advance\pfrontx by 5000
\global\advance\pfronty by -500
\put (\pfrontx,\pfronty){$+$}
\global\advance\pfronty by 500
\global\advance\pfrontx by 5000

\drawline\fermion[\NW\REG](\pfrontx,\pfronty)[3000]
\drawline\fermion[\SW\REG](\pfrontx,\pfronty)[3000]
\drawline\scalar[\E\REG](\pfrontx,\pfronty)[3]
\put(\pmidx,18300){\footnotesize $H$}
\drawline\fermion[\NE\REG](\pbackx,\pbacky)[3000]
\drawline\fermion[\SE\REG](\pfrontx,\pfronty)[3000]

\put (26700,20000){$\pi$}
\put (26700,15000){$\pi$}
\put (38000,20000){$\pi$}
\put (38000,15000){$\pi$}

\drawline\fermion[\SE\REG](7500,13000)[6000]
\global\advance\pmidx by 150
\put (\pmidx,\pmidy){\circle*{700}}
\global\advance\pmidx by -150
\drawline\fermion[\NE\REG](\pmidx,\pmidy)[3000]
\drawline\fermion[\SW\REG](\pfrontx,\pfronty)[3000]

\put (4000,\pfronty){\small b)}

\put (6700,13000){$\pi$}
\put (6700,8000){$\pi$}
\put (12300,13000){$\pi$}
\put (12300,8000){$\pi$}

\global\advance\pfrontx by 5000
\global\advance\pfronty by -500

\put (\pfrontx,\pfronty){$+$}
\global\advance\pfronty by 500

\global\advance\pfrontx by 5000

\drawline\fermion[\SW\REG](\pfrontx,\pfronty)[3000]
\drawline\fermion[\NW\REG](\pfrontx,\pfronty)[3000]
\put (\pfrontx,\pfronty){\circle*{700}}
\global\advance\pfrontx by 1000
\put (\pfrontx,\pfronty){\circle{2000}}
\global\advance\pfrontx by 1000
\put (\pfrontx,\pfronty){\circle*{700}}
\drawline\fermion[\NE\REG](\pfrontx,\pfronty)[3000]
\drawline\fermion[\SE\REG](\pfrontx,\pfronty)[3000]

\put (16700,13000){$\pi$}
\put (16700,8000){$\pi$}
\put (24300,13000){$\pi$}
\put (24300,8000){$\pi$}

\global\advance\pfrontx by 5000
\global\advance\pfronty by -500
\put (\pfrontx,\pfronty){$+$}
\global\advance\pfronty by 500
\global\advance\pfrontx by 5000

\drawline\fermion[\SW\REG](\pfrontx,\pfronty)[3000]
\drawline\fermion[\NW\REG](\pfrontx,\pfronty)[3000]
\put (\pfrontx,\pfronty){\circle*{700}}
\global\advance\pfrontx by 1000
\put (\pfrontx,\pfronty){\circle{2000}}
\global\advance\pfrontx by 1000
\put (\pfrontx,\pfronty){\circle*{700}}
\global\advance\pfrontx by 1000
\put (\pfrontx,\pfronty){\circle{2000}}
\global\advance\pfrontx by 1000
\put (\pfrontx,\pfronty){\circle*{700}}
\drawline\fermion[\NE\REG](\pfrontx,\pfronty)[3000]
\drawline\fermion[\SE\REG](\pfrontx,\pfronty)[3000]

\put (28800,13000){$\pi$}
\put (28800,8000){$\pi$}
\put (38000,13000){$\pi$}
\put (38000,8000){$\pi$}

\global\advance\pfrontx by 4000
\global\advance\pfronty by -250
\put (\pfrontx,\pfronty){$+ \ \dots$}

\put(4000,5000){\small c)}

\drawline\scalar[\E\REG](6500,5000)[4]

\global\advance \pfrontx by 300
\put(\pfrontx,5300){\footnotesize $H$}
\global\advance \pbackx by -300
\put(\pbackx,5300){\footnotesize $H$}
\global\advance \pbackx by 300
\put(\pmidx,\pmidy){\circle*{1200}}

\global\advance \pbackx by 2000
\global\advance\pbacky by -250
\put (\pbackx,\pbacky){$=$}
\global\advance\pbacky by 250
\global\advance\pbackx by 2000

\drawline\scalar[\E\REG](\pbackx,\pbacky)[3]

\put(\pmidx,5300){\footnotesize $H$}
 
\global\advance\pbackx by 2000
\global\advance\pbacky by -250
\put (\pbackx,\pbacky){$+$}
\global\advance\pbacky by 250
\global\advance\pbackx by 2000

\drawline\scalar[\E\REG](\pbackx,\pbacky)[2]
\global\advance \pfrontx by 300
\put(\pfrontx,5300){\footnotesize $H$}
\global\advance\pbackx by 1100
\put(\pbackx,\pbacky){\circle{2000}}
\global\advance\pbackx by 1000
\drawline\scalar[\E\REG](\pbackx,\pbacky)[2]
\global\advance \pbackx by -300
\put(\pbackx,5300){\footnotesize $H$}

\put(9000,1250){$+$}
\drawline\scalar[\E\REG](11000,1500)[2]
\global\advance \pfrontx by 300
\put(\pfrontx,1700){\footnotesize $H$}
\global\advance\pbackx by 1000
\put(\pbackx,\pbacky){\circle{2000}}
\global\advance\pbackx by 1000
\put(\pbackx,\pbacky){\circle*{700}}
\global\advance\pbackx by 1000
\put(\pbackx,\pbacky){\circle{2000}}
\global\advance\pbackx by 1000
\drawline\scalar[\E\REG](\pbackx,\pbacky)[2]
\global\advance \pbackx by -300
\put(\pbackx,1700){\footnotesize $H$}

\global\advance\pbackx by 2000
\global\advance\pbacky by -250
\put (\pbackx,\pbacky){$+$}
\global\advance\pbacky by 250
\global\advance\pbackx by 2000

\drawline\scalar[\E\REG](\pbackx,\pbacky)[2]
\global\advance \pfrontx by 300
\put(\pfrontx,1700){\footnotesize $H$}
\global\advance\pbackx by 1000
\put(\pbackx,\pbacky){\circle{2000}}
\global\advance\pbackx by 1000
\put(\pbackx,\pbacky){\circle*{700}}
\global\advance\pbackx by 1000
\put(\pbackx,\pbacky){\circle{2000}}
\global\advance\pbackx by 1000
\put(\pbackx,\pbacky){\circle*{700}}
\global\advance\pbackx by 1000
\put(\pbackx,\pbacky){\circle{2000}}
\global\advance\pbackx by 1000
\drawline\scalar[\E\REG](\pbackx,\pbacky)[2]
\global\advance \pbackx by -300
\put(\pbackx,1700){\footnotesize $H$}

\global\advance\pbackx by 2000
\global\advance\pbacky by -250
\put (\pbackx,\pbacky){$+ \ \dots$}

\end{picture}

\leftskip 1cm
\rightskip 1cm
{\footnotesize {\bf Figure 1: }
Diagrams contributing to: a) The tree level Goldstone boson scattering amplitude.
b) The leading order in the $1/N$ expansion for the same process. 
c) The Higgs propagator at
leading order in the $1/N$ expansion.}

\leftskip 0cm
\rightskip 0cm

\end{figure}

The tree level contributions to the $A$ function $(A_0)$ are obtained from the
diagrams in Fig.1.a
\begin{equation}
A_0(s)=-2\lambda\left(1+\frac{M^2_H}{s-M^2_H}    \right)=
\frac{s}{NF^2}\frac{1}{1-s/M_H^2} 
\end{equation}
and therefore they only depend on $s$. In the large $N$ limit, the relevant
diagrams are those shown in Fig.1.b, which are known as bubble diagrams.
Each of the loops contributes with the same factor and the sum
of all those diagrams can be seen as a geometric series which amounts to
\begin{equation}
A(s)=
\frac{s}{NF^2}\frac{1}{1-s/M_H^2+sI(s)/2F^2} 
\label{AlargeN}
\end{equation}
where the divergent one-loop integral $I(s)$ can be calculated using
dimensional regularization. The result is
\begin{equation}
I(s)=\frac{-1}{(4\pi)^2}\left(N_{\epsilon}+2-\log\frac{-s}{\mu}  \right) ,
 \end{equation}
where as usual
\begin{equation}
N_{\epsilon}= \frac{2}{\epsilon}+\log 4\pi-\gamma_E ,
 \end{equation}
and $\mu$ is an arbitrary renormalization scale. Thus, in the large $N$ limit
the $A$ function only depends on $s$. The $1/\epsilon$ divergencies appearing 
in $I(s)$ can be absorbed in the renormalized Higgs mass $M_R^2$ which can be defined
as 
\begin{equation}
\frac{1}{M_R^2}\equiv\frac{1}{M_H^2}+\frac{N_{\epsilon}+2}{2(4\pi)^2F^2}
\end{equation}
so that we find
\begin{equation}
A(s)=
\frac{s}{NF^2}\frac{1}{1-\frac{s}{M_R^2}+\frac{s}{2(4\pi)^2F^2}\log\frac{-s}
{\mu^2}} 
\end{equation}
In this approach the Higgs mass is the only parameter that needs 
renormalization and in  particular there is no wave function renormalization.
Thus the above amplitude is an observable and $\mu$ independent quantity. This
fact can be  used to find the dependence of the renormalized Higgs mass $M_R$
on the renormalization scale $\mu$ which turns out to be
\begin{equation}
M_R^2(\mu)=
\frac{M_R^2(\mu_0)}{1-\frac{M_R^2(\mu_0)}{2(4\pi)^2F^2}\log\frac{\mu^2}
{\mu^2_0}} 
\end{equation}
The renormalized coupling $\lambda_R$ can be defined in order to keep the tree level
relation $M_R^2=2\lambda_RNF^2$ and then its running can be easily obtained from
the above evolution equation. In practice it is useful to
introduce the mass parameter $M^2$ defined by the equation
\begin{equation}
M^2=M_R^2(M^2)
\label{masspara}
\end{equation}
and then
\begin{equation}
M_R^2(\mu)=
\frac{M^2}{1-\frac{M^2}{2(4\pi)^2F^2}\log\frac{\mu^2}{M^2}} 
\label{MR}
\end{equation}
so that
\begin{equation}
\lambda_R(\mu)=
\frac{\lambda(M)}{1-\frac{N\lambda(M)}{(4\pi)^2}\log\frac{\mu^2}{M^2}} 
\end{equation}
From this formula we can obtain the position $\Lambda$ of the Landau pole  in
this approximation which is given by
\begin{equation}
\Lambda^2=M^2e^{\frac{(4 \pi)^2}{N\lambda(M)}} 
\end{equation}

Therefore, for $g=g'=0$ the mass parameter is the only free parameter
of the model and all the observables can be obtained in terms of it. However,
this  mass should not be confused with the physical Higgs mass. The physical
mass is the mass of  the resonance appearing in the scattering channel with
the same quantum numbers as the Higgs particle.

In the real world,
where $N=3$, the coset space is $O(4)/O(3)=SU(2)_L\times
SU(2)_R/SU(2)_{L+R}$ and thus the interactions are $SU(2)_{L+R}$ symmetric 
(weak isospin group). Hence there are three Goldstone bosons  and the scattering
channels can be labelled by the third component of the isospin which can take
the values $I=0,1,2$. For an arbitrary $N$ it
is still possible to define the appropriate generalization of the above 
mentioned channels which  are then given by \cite{Dugan}
\begin{eqnarray}
T_0(s,t,u) & = & NA(s,t,u)+A(t,s,u)+A(u,t,s)   \nonumber  \\
T_1(s,t,u) & = & A(t,s,u)-A(u,t,s)   \nonumber  \\
T_2(s,t,u) & = & A(t,s,u)+A(u,t,s)
\end{eqnarray}
Let us now recall that in Eq.\ref{AlargeN} we had found that $A(s,t,u)\simeq 
A(s)\sim\Od(1/N)$
and therefore
\begin{equation}
T_0=NA(s)+\Od(1/N)
\end{equation}
is the only non zero isospin channel in the large $N$ limit. 
Fortunately, that is precisely the channel where the Higgs 
would appear.
Customarily the
amplitudes are also projected in definite total angular
momentum states, leading to partial waves $t_{IJ}$. 
It is also obvious that in this case only the $t_{00}$
survives since $T_0$ only depends on $s$. Indeed
\begin{equation}
t_{00}(s)=\frac{s}{32\pi F^2}\left(
1-\frac{s}{M^2}+\frac{s}{2(4\pi)^2F^2}\log\frac{-s}{M^2}
\right)^{-1}+\Od\left(\frac{1}{N}\right)
\label{t00}
\label{pawa}
\end{equation}
This partial wave has some properties which make the large $N$ limit a
sensible approximation to Higgs physics. First, at low energies we find
\begin{equation}
t_{00}(s)\simeq \frac{s}{32\pi F^2}
\end{equation}
in agreement with the Weinberg low-energy theorems. Second, this partial
wave has the correct unitarity cut along the positive real axis of the $s$
variable. Indeed, it can be easily
checked that for physical $s$ values, which are located
 right on the unitary cut where $\log
(-s)=\log s -i\pi$, we have 
\begin{equation}
\Ima \; t_{00}= \mid t_{00} \mid^2+\Od(1/N)
\end{equation}
which is the elastic unitarity condition. 

\begin{figure}

\vspace{-0.8cm}

\leftskip -2.2cm
\begin{center}
\mbox{\epsfysize=6.8cm\epsffile{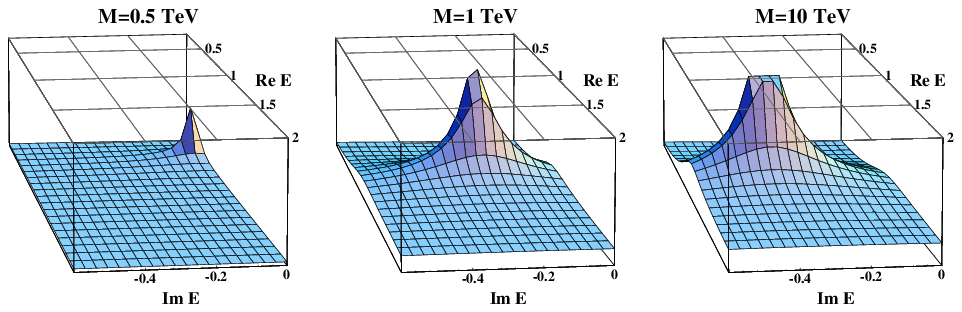}}

\leftskip 1cm
\rightskip 1cm

\vskip -.5cm

{\footnotesize {\bf Figure 2: }
Evolution of the position of the $\vert t_{00}\vert$
pole in the $E=\sqrt{s}$ complex plane.
We display the lower half of the
second Riemann sheet as a function of the $M$ parameter. Notice how
the distance to the real axis grows with $M$, whereas the real part of the
position remains bounded. The scale is the same for the three figures.}

\leftskip 0cm
\rightskip 0cm
\end{center}
\end{figure}

\vskip .5cm

Finally, we want to remark that
it is possible to find numerically
that the partial wave in Eq.\ref{pawa} has a pole in
the second Riemann sheet. This pole can be understood as the physical Higgs
resonance. In Fig.2 it is shown the position of this pole in the complex
plane for different $M$ values. 

For low $M$ values the physical Higgs
resonance is narrow and the standard Breit-Wigner description of the
resonance can be safely applied. 
Then the physical mass is just given by $M$ whereas the width is
\begin{equation}
\Gamma=\frac{M^3}{32 \pi F^2}
\end{equation}
which is the tree level result. However, when $M$ increases, the Higgs
resonance becomes broader and broader. The pole migrates down in the complex
plane  and the Breit-Wigner description cannot be used
any more. However, the real part of the pole position remains bounded even
for very large $M$ as can be seen in Fig.4. This feature is
usually called "saturation" and it has also
been observed in other non-perturbative approaches to the Higgs dynamics. In particular 
this behaviour was obtained using the $N/D$ method in \cite{LQT} and \cite{N/D}, using
the Pad\'e approximants in \cite{Pade} and using the large $N$ limit in \cite{LargeN}.

\leftskip -2cm
\begin{center}
\mbox{\epsfysize=8cm\epsffile{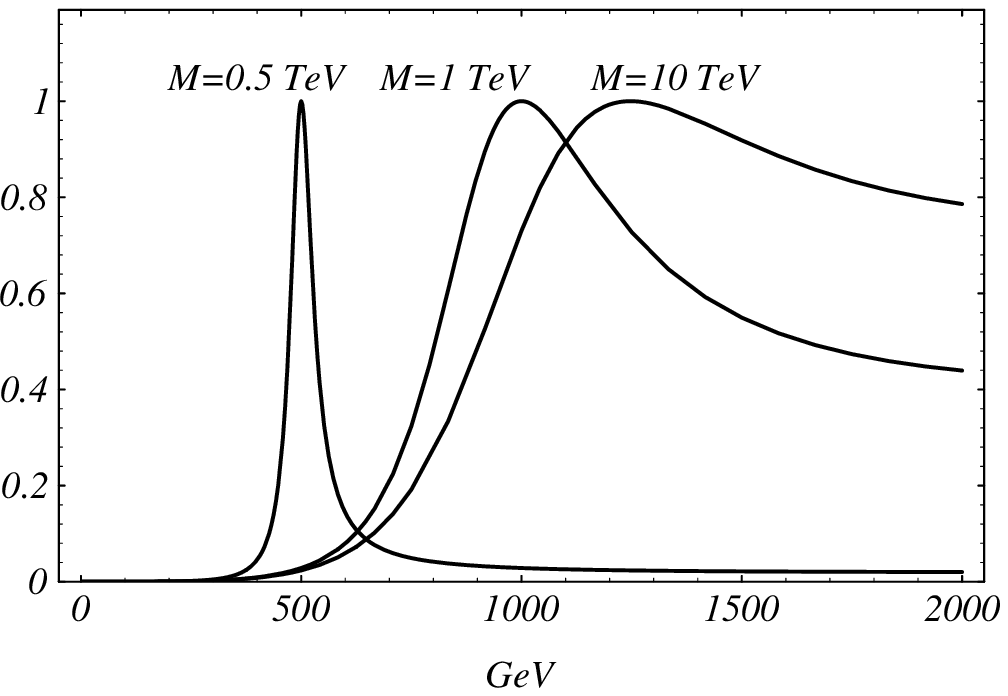}}
\end{center}

\vspace{-.5cm}

\leftskip 1cm
\rightskip 1cm

{\footnotesize {\bf Figure 3: }
$\vert t_{00} \vert^2$ versus $\sqrt{s}$ for
different values of the Higgs mass parameter $M$ as defined in Eq.\ref{masspara}. 
Even for values as large as $M=10\mbox{TeV}$, the position
of the resonance is not higher than $1.5 \mbox{TeV}$. }

\leftskip 0.cm
\rightskip 0.cm

\section{Gauge boson scattering and the Equivalence Theorem}

We have already stated that our aim in this work is to study the large $N$
limit of the  Higgs sector including the electroweak gauge
bosons. More precisely we are considering the $N\rightarrow\infty$ limit but
keeping $Ng^2$ and $Ng'^2$ constant. We will see that such an approach to 
the gauged Higgs sector turns out to be very useful since it provides a sensible
description of gauge boson interactions that still allows easy calculations.

 In the following we will concentrate in the elastic scattering
process $VV \rightarrow VV$ where $V=W^{\pm },Z^0$. In order to obtain the
leading contribution in the approximation defined above, the first observation
is that the diagrams at tree level
are $\Od(g^2)$ (or $\Od(g'^2)$).
Due to the particular way in which the large $N$ limit has been defined, those
graphs are $\Od(1/N)$ too. To find the
complete set of diagrams contributing to the large $N$ leading order, 
we have to include into the tree diagrams any possible
internal loop without increasing their $g^2,g'^2$ or $1/N$ power dimensions.
It is fairly simple to see that that cannot be accomplished
with gauge boson loops.
Concerning the scalars, the relevant observation is that gauge bosons are only coupled
to the three first $\pi^a$, whereas the Higgs interacts with all them. 
Thus, the only $\pi$ loops appearing in the large $N$ limit are those coupled to
the Higgs field.

The main effect of those $\pi$ loops is to contribute to the Higgs
propagator as it is shown in Fig.1.c. Note that, as far as we are working in
the Landau gauge, where all the $\pi$ fields are massless, many other
possible $\pi$ loop diagrams vanish, since 
they are proportional to $\int d^{4-\epsilon}q/q^2 $ which
is zero when using dimensional regularization.

It is not very difficult to calculate the diagrams in Fig.1.c. Using the
renormalization prescription of the previous section for the renormalized Higgs
mass, we find the Higgs propagator
\begin{equation}
D(q^2)=\frac{1}{q^2-M_R(-q^2)}+\Od\left(\frac{1}{N}\right)
\label{Nprop}
\end{equation}
where $M_R(-q^2)$ is defined in Eq.\ref{MR}. It is obvious that this $D(q^2)$ has
exactly the same pole in the second Riemann sheet than the $t_{00}$ partial
wave amplitude in Eq.\ref{t00}, which corresponds to the physical Higgs resonance. At the
same time, for small $M$, we find $M_H(-q^2)\rightarrow M$ and thus we
recover the standard perturbative (tree level) behavior of the Higgs
resonance whose width would then be given by Eq.23. Therefore the above
propagator describes properly the Higgs resonances both in the perturbative (light Higgs)
and the non-perturbative regime (heavy Higgs).

\begin{figure}
\small
\begin{picture}(45000,22000)

\THICKLINES
\bigphotons

\put (2000,17000){{\normalsize a)}}

\drawline\photon[\E\REG](6500,22000)[4]
\drawline\photon[\SE\CURLY](\pbackx,\pbacky)[6]
\drawline\photon[\S\REG](\pfrontx,\pfronty)[4]
\drawline\photon[\W\REG](\pbackx,\pbacky)[4]
\drawline\photon[\NE\FLIPPEDCURLY](\pfrontx,\pfronty)[2]
\global\advance\pbackx by 900
\global\advance\pbacky by 900
\drawline\photon[\NE\FLIPPEDCURLY](\pbackx,\pbacky)[3]

\put (4500,22000){$W^+$}
\put (4500,18000){$W^-$}
\put (15500,22000){$Z^0$}
\put (15500,18000){$Z^0$}

\drawline\photon[\E\REG](21500,22000)[4]
\drawline\fermion[\S\REG](\pbackx,\pbacky)[4000]
\drawline\photon[\W\REG](\pbackx,\pbacky)[4]
\drawline\fermion[\N\REG](\pfrontx,\pfronty)[4000]
\drawline\photon[\SE\CURLY](\pbackx,\pbacky)[6]
\drawline\fermion[\S\REG](\pfrontx,\pfronty)[4000]
\drawline\photon[\NE\FLIPPEDCURLY](\pbackx,\pbacky)[2]
\global\advance\pbackx by  900
\global\advance\pbacky by  900
\drawline\photon[\NE\FLIPPEDCURLY](\pbackx,\pbacky)[3]

\put (19500,22000){$W^+$}
\put (19500,18000){$W^-$}
\put (30200,22000){$Z^0$}
\put (30200,18000){$Z^0$}

\drawline\photon[\SE\CURLY](33800,22000)[3]
\drawline\photon[\SW\CURLY](\pbackx,\pbacky)[3]
\drawline\scalar[\E\REG](\pfrontx,\pfronty)[3]
\put(\pmidx,\pmidy){\circle*{1200}}
\drawline\photon[\NE\CURLY](\pbackx,\pbacky)[3]
\drawline\photon[\SE\FLIPPEDCURLY](\pfrontx,\pfronty)[3]

\put (32600,22000){$W^+$}
\put (32600,18000){$W^-$}
\put (44300,22000){$Z^0$}
\put (44300,18000){$Z^0$}


\drawline\photon[\E\REG](6500,14000)[8]
\drawline\photon[\S\REG](\pmidx,\pmidy)[4]
\drawline\photon[\E\REG](\pbackx,\pbacky)[4]
\drawline\photon[\W\REG](\pfrontx,\pfronty)[4]

\put (4500,14000){$W^+$}
\put (4500,10000){$W^-$}
\put (15500,14000){$Z^0$}
\put (15500,10000){$Z^0$}

\drawline\photon[\E\REG](21500,14000)[8]
\drawline\fermion[\S\REG](\pmidx,\pmidy)[4000]
\drawline\photon[\E\REG](\pbackx,\pbacky)[4]
\drawline\photon[\W\REG](\pfrontx,\pfronty)[4]

\put (19500,14000){$W^+$}
\put (19500,10000){$W^-$}
\put (30200,14000){$Z^0$}
\put (30200,10000){$Z^0$}

\drawline\photon[\SE\CURLY](37500,14000)[3]
\drawline\photon[\SE\FLIPPEDCURLY](\pbackx,\pbacky)[3]
\drawline\photon[\NE\CURLY](\pfrontx,\pfronty)[3]
\drawline\photon[\SW\CURLY](\pfrontx,\pfronty)[3]

\put (35500,14000){$W^+$}
\put (35500,10000){$W^-$}
\put (42200,14000){$Z^0$}
\put (42200,10000){$Z^0$}


\put (2000,3500){{\normalsize b)}}

\drawline\fermion[\E\REG](10500,5500)[8000]
\drawline\photon[\S\REG](\pmidx,\pmidy)[4]
\drawline\fermion[\E\REG](\pbackx,\pbacky)[4000]
\drawline\fermion[\W\REG](\pfrontx,\pfronty)[4000]

\put (8500,5500){$\pi^+$}
\put (8500,1500){$\pi^-$}
\put (19500,5500){$\pi^0$}
\put (19500,1500){$\pi^0$}

\drawline\fermion[\E\REG](26500,5500)[4000]
\drawline\fermion[\SE\REG](\pbackx,\pbacky)[5500]
\drawline\photon[\S\REG](\pfrontx,\pfronty)[4]
\drawline\fermion[\W\REG](\pbackx,\pbacky)[4000]
\drawline\fermion[\NE\REG](\pfrontx,\pfronty)[2000]
\global\advance\pbackx by 900
\global\advance\pbacky by 900
\drawline\fermion[\NE\REG](\pbackx,\pbacky)[2500]

\put (24500,5500){$\pi^+$}
\put (24500,1500){$\pi^-$}
\put (35500,5500){$\pi^0$}
\put (35500,1500){$\pi^0$}

\end{picture}

\leftskip 1cm
\rightskip 1cm
{\footnotesize {\bf Figure 4: }
a) Diagrams contributing to the $W^+W^- \rightarrow
Z^0Z^0$ process at leading order in the $1/N$ expansion.
b) Tree level diagrams contributing to the $\pi^+\pi^- \rightarrow
 \pi^0\pi^0$ amplitude containing an internal gauge boson line.}

\leftskip 0.cm
\rightskip 0.cm
\normalsize
\end{figure}

The most relevant consequence of the previous discussion is
that the $VV \rightarrow VV$ leading 
diagrams are just those at tree level, but using the above Higgs
propagator instead of that calculated at tree level. For example, the 
contributions to $W^+W^- \rightarrow Z^0Z^0$ can be found in Fig.4.a. Thus in
this limit the calculations are not much more difficult than at tree level.
However, the unitarity properties of the large $N$ amplitudes are
greatly improved and the Higgs mass and width is properly described in a way
which is compatible with other non-perturbative approaches.

An important test for the consistency of the approximation is
provided by the Equivalence Theorem (ET). This theorem states that the
$S$-matrix elements of longitudinal electroweak gauge boson are the same as
those of their associated
would-be Goldstone bosons, up to $\Od(m/E)$ corrections,
where $m=m_W,m_Z$ and $E$ is the typical C.M. energy of the process. Thus,
on the one hand, at
high energies the scattering of longitudinal gauge bosons provides
information about the Higgs sector of the SM. On the other hand, the ET
can be used to calculate the longitudinal gauge boson scattering at high
energies in terms of scalars, which are much easier to handle.
In fact most of the calculations performed for the LHC
until now have used this theorem. 

In the approach followed here we are
including explicitly the gauge degrees of freedom and therefore we do 
{\em not} need
to use the ET at all. As a consequence, our approach will be more
reliable at lower energies than if we had used the ET, which is
neglecting ${\cal O}(m/E)$ terms.
Nevertheless, the theorem can be useful as a
tool to check our results. For example, it relates 
at high energies the $W^+W^-
\rightarrow Z^0Z^0$ and the $\pi^+\pi^- \rightarrow \pi^0\pi^0$ $S$-matrix
elements. At this moment a few comments are in order. First
the $S$-matrix elements in both sides of the theorem
can be expanded in terms of $1/N$ and thus it
should apply order by order in $1/N$. In this
work we are considering the $N\rightarrow\infty$ with $\lambda N$,
$g^2N$ and $g'^2N$ constant. 
In particular that means that for the $\pi^+\pi^- \rightarrow
\pi^0\pi^0$ process one has to include, at leading order, the diagrams
in Fig.4.b in
addition to those in Fig.1. This is because in the previous section
our model had not been gauged yet, but once it is gauged
the new diagrams which are
${\cal O}(g^2)$ are also ${\cal O}(1/N)$ and they should not be forgotten.
these new diagrams are $\Od
(g^2)$ whereas those considered in the previous sections are simply $\Od(1/N)$. 

Thus the leading order for this amplitude reads
\begin{eqnarray}
 T(\pi^+\pi^- \rightarrow\pi^0\pi^0) & = &
\frac{s}{s-M^2_R(-s)}\frac{1}{4(1-x^2)-4\frac{m_W^2}{E^2}+\frac{m_W^4}{E^4}}
\nonumber \\
&\cdot& \left[ -4\frac{M_R^2(-s)}{v^2}(1-x^2)+2g^2(3+x^2)
-2\frac{M_R^2(-s)}{v^2}\frac{m_W^2}{E^2}(5+x^2) \right.
\nonumber \\
&& \left.+ 12\frac{m_W^2}{v^2}\frac{m_W^2}{E^2}    
-4\frac{M_R^2(-s)}{v^2}\frac{m_W^4}{E^4} \right] 
\end{eqnarray}
 where $s=4E^2$, $E$ is the $\pi$ energy, $\theta$ is the scattering angle and $M_R(-s)$ can
be obtained from Eq.15. Note that, as far as $-s$ is negative, $M_R(-s)$ produces the imaginary
part and the cut for the above amplitude required for unitarity.

After a lengthy but straightforward calculation using the Feynman
rules coming from the lagrangian in Eq.2
(plus the standard gauge fixing and Faddeev-Popov
terms) and projecting out the longitudinal components, 
we arrive to the following result for the $W^+_LW^-_L \rightarrow Z_LZ_L$ scattering amplitude:
\begin{eqnarray}
T(W^+_LW^-_L \rightarrow Z_LZ_L)  & = &
\frac{s}{s-M^2_R(-s)}\frac{1}{4(1-x^2)-4\frac{m_W^2}{E^2}(1-2x^2)
+\frac{m_W^4}{E^4}(1-4x^2)} \nonumber \\
&\cdot& \left[ - 4\frac{M_R^2(-s)}{v^2}(1-x^2)+2g^2(3+x^2)+
2\frac{M_R^2(-s)}{v^2}\frac{m_W^2}{E^2}(1-7x^2) \right. \nonumber \\
&&+4\frac{m_W^2}{v^2}\frac{m_W^2}{E^2}(-14+5x^2) +  
8\frac{M_R^2(-s)}{v^2}\frac{m_W^4}{E^4}(1+x^2)+
8\frac{m_W^2}{v^2}\frac{m_W^4}{E^4}(3+x^2)\nonumber \\
&&\left. -4\frac{M_R^2(-s)}{v^2}\frac{m_W^6}{E^6}(1+2x^2) 
- \frac{m_W^2}{v^2}\frac{m_W^6}{E^6}(1-4x^2) \right]
\end{eqnarray}

As expected, it can be easily checked that
 these two amplitudes satisfy the ET. One potential problem 
that could appear when using the ET
comes from the different
renormalization of the gauge boson and $\pi$ wave functions \cite{RET}. 
Fortunately, at leading
order our $1/N$ expansion does not need wave function renormalization and
the ET can be safely applied.

In order to illustrate the above discussion and to check 
our computational  methods
we have displayed in Fig.5 the scattering cross section of
$W_L^+W_L^- \rightarrow Z_L^0Z_L^0$ versus that of
$\pi^+\pi^- \rightarrow \pi^0\pi^0$. The former is 
represented by a continuous line
whereas the latter has been drawn discontinuously. Notice that
to all means and purposes they overlap {\em at high energies} ($E>1.2$ TeV).

From Fig.5 we can observe that either with or without the ET,
the large $N$ approximation is able to
reproduce a well shaped Higgs resonant behaviour and very good high
energy properties.  
The small numerical differences up to almost 1.2 TeV are simply
due to the fact that the ET is neglecting the ${\cal O}(m/E)$ 
contributions.
Thus we can summarize these two last sections by saying that
the large $N$ meets in a very simple way all  the
known theoretical constraints to the SM Higgs sector, like 
the low-energy theorems,
unitarity, the saturation property and the ET.

\vspace{-1.2cm}

\leftskip -2cm
\begin{center}
\mbox{\epsfysize=8cm\epsffile{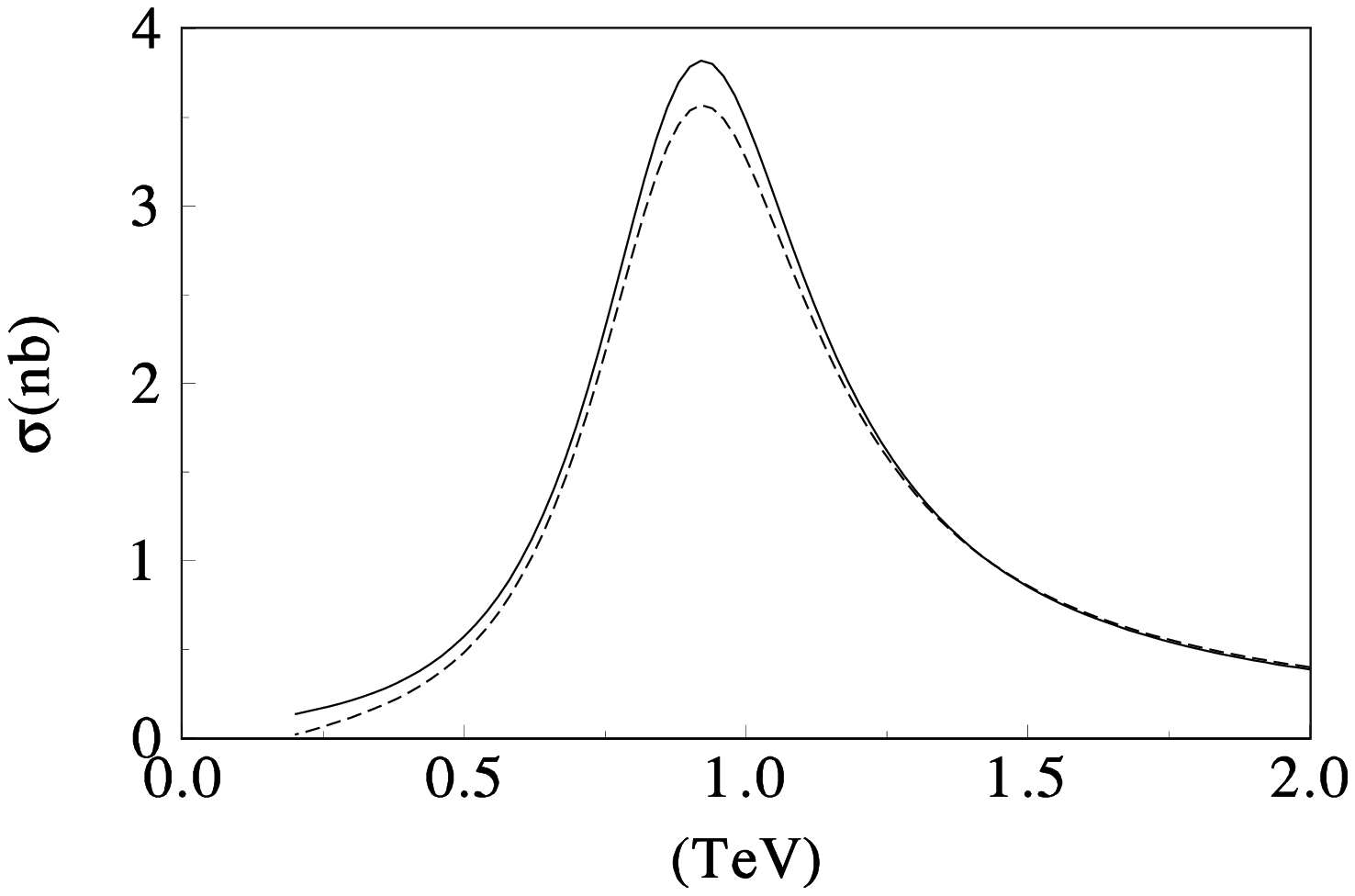}}
\end{center}

\vspace{-.2cm}

\leftskip 1cm
\rightskip 1cm
{\footnotesize {\bf Figure 5: }
Comparison of the total $W^+W^- \rightarrow
 Z^0Z^0$ cross-section
at different $\sqrt{s}$, for $\mid \cos \theta \mid <0.8$, 
calculated with our large $N$ approach,
either with (dashed line)
or without the ET (continuous line).}

\leftskip 0.cm
\rightskip 0.cm

\section{Numerical results for the LHC}

The main practical application of the approach described above is of course the
description of the LHC phenomenology. For this reason it will be used in this
section to obtain predictions in terms of the
renormalized Higgs mass under the hypothesis 
that the MSM provides the right model
for the electroweak symmetry breaking. In particular we will concentrate on 
$Z^0Z^0$ pair production, since this final state is the most sensible to the
Higgs resonance properties and at the same time gives rise to a very clear
experimental signature.

We consider both final gauge bosons decaying into the cleanest 
leptonic channels: $Z^0 \rightarrow e^+e^-,\;\mu^+ \mu^-$. 
Indeed, we have obtained the number of these events as the
total number of $Z^0Z^0$ pairs times the branching ratio $0.0044$. 
We have computed the total $Z^0Z^0$ number of events at the LHC with 
the help of the Monte-Carlo VEGAS code \cite{Vegas}. 
In order to relate
the subprocesses cross sections to the $pp$ initial state,
we have used the effective $W$ approximation  
\cite{Daw} (which is based on the Weizsaker-Williams approximation
\cite{WW}) and the MRSD \cite{MRS} proton structure functions, which
are in good agreement with recent experimental results at HERA.

The different subprocesses
contributing to $Z^0Z^0$ production that we have evaluated are  
\begin{eqnarray}
Z^0Z^0 &  \rightarrow & Z^0Z^0  \nonumber  \\      
W^+W^-   &  \rightarrow & Z^0Z^0  \nonumber \\
q \bar{q} & \rightarrow & Z^0Z^0   \nonumber \\ 
gg   & \rightarrow  &  Z^0Z^0
\label{eq:procesos}
\end{eqnarray}
All these channels have been calculated using the MSM Feynman rules within the  
large $N$  limit, which modifies the Higgs boson mass and width
according to our previous discussion. Consequently we have used the Higgs
propagator given in Eq.\ref{Nprop}, so that $M$ remains as a free parameter. We
have evaluated most of the cross sections  shown in Eq.\ref{eq:procesos}  at tree
level, although  gluon-gluon fusion is calculated to one-loop \cite{Glo}, since
it occurs via quark loops. As a consequence
this cross section is quite sensitive to the top quark mass,
that has been set to  $m_t=180$ GeV. 
 
In order to compute the total number of events of the subprocesses
in Eq.\ref{eq:procesos} we have
 set the following expected values for the LHC parameters:
the $pp$ center of mass 
energy, $\sqrt{s}=14$ TeV and an integrated luminosity 
$L=3\times10^{5}\mbox{pb}^{-1}$. In addition, 
we choose the following kinematical   
cuts on the maximum $Z^0Z^0$ invariant mass ($\sqrt{\hat{s}}^{\rm max}=5$ TeV), the
minimum  transverse momentum ($p_{tZ}^{\rm
min}=300$ GeV) and the maximum rapidity $y_{Z}^{\rm max}=2$. 
Finally, in order to test the dependence on the renormalized mass
parameter,
we have chosen different input values for $M$: 
$100$, $500$  and $1000$ GeV as defined in Eq.\ref{masspara},
which cover a wide variety of regimes, from weak to strongly interacting.
The results are displayed in Table 1.

\begin{table}
\vspace{0.7cm}
\centering
\begin{tabular}{|c|c|c|c|} \hline
\rule[-3mm]{0mm}{8mm} & $M=100 \;$GeV & $M=500 \;$GeV & $M=1 \;$TeV  \\  \hline
$Z^0Z^0 \rightarrow Z^0Z^0$ & 0.09 & 2.58 & 8.95  \\  \hline
$W^+W^- \rightarrow Z^0Z^0$  & 21.23 & 23.39 & 46.32  \\  \hline
$q\bar{q} \rightarrow Z^0Z^0$ & \multicolumn{3}{c|}{ \hspace{.5cm}53.83 }\\ \hline
$gg \rightarrow Z^0Z^0$ & \multicolumn{3}{c|}{ \hspace{.5cm}13.42 } \\ \hline\hline

\rule[-3mm]{0mm}{8mm}
$Z^0Z^0+W^+W^- \rightarrow Z^0Z^0$ & 21.33 & 25.97 & 55.27  \\  \hline
\rule[-3mm]{0mm}{8mm}
All$\rightarrow Z^0Z^0$ & 88.57  & 93.21 & 122.52  \\  \hline
\end{tabular}

\vskip .5cm
\leftskip .8cm
\rightskip .8cm

{\footnotesize {\bf Table 1:}
Total number of $Z^0Z^0$ events at LHC decaying to 
the cleanest leptonic decays ($e,\mu$), in the 
large $N$ limit of the SM.
We have set the following kinematical cuts
on the final $Z^0$ bosons: $\sqrt{\hat{s}}^{\rm max}=5$ TeV, $p_{tZ}^{\rm
min}=300$ GeV, $y_{Z}^{\rm max}=2$. To illustrate the effect of changing
the renormalized Higgs mass $M$ in Eq.\ref{MR},  
we have chosen three typical values. 
The contributions from different initial subprocesses
are shown explicitly, although    
those events coming from other gauge boson pairs are listed together. The
top quark mass has been fixed to $m_t=180$ GeV.}

\leftskip 0.cm
\rightskip 0.cm

\end{table}

\section{Conclusions}
\hspace*{12pt}

We have studied the main properties of the Standard Model Higgs sector in the large $N$
limit, i.e. for a large number of would-be Goldstone bosons, including    the $SU(2)_L
\times U_Y(1)$ interactions,  keeping $N\lambda$, $Ng^2$ and $Ng'^2$ constant. By using
this approximation we have confirmed
the expected behaviour from other non-perturbative approaches, 
both in the weak and the strong interaction regime.
In particular the Higgs mass saturation property. In addition we have been able to
give a proper description of the Higgs resonance as a pole in the second Riemann sheet 
of the $I=J=0$ channel, thus having a well defined width. The corresponding
partial wave has very good unitarity properties and it is compatible with the  low-energy
theorems. Furthermore, the explicit introduction of gauge fields
as well as the simplicity to implement this approach allow us,
in contrast to most of the previous approaches, to obtain the
$W^+$, $W^-$ and $Z$ scattering  amplitudes
by means of very simple calculations, even
 {\em without} the help of the Equivalence Theorem,
which nevertheless has been used to check our results.
As an illustration we have applied the large $N$
approximation to estimate the number of $Z^0Z^0$ events with the cleanest
signature at the LHC, including all relevant backgrounds. The results can be found in
the table. As it can be seen there, the sensibility of the number of events to the Higgs
mass parameter is not very large. However, it could by considerably increased 
with jet
tagging, which could help to separate the more interesting
pure fusion events from the background.

We have therefore shown how the large $N$, despite its simplicity
(only the propagator has to be 
changed), yields a consistent description of the 
Higgs sector non-perturbative problems, thus
improving previous approaches used
to obtain predictions for the LHC. 

\vskip 1.0cm
\section*{Acknowledgments} 

This work has been supported in part by the Ministerio de Educaci\'on y
Ciencia (Spain) (CICYT AEN95-1285-E) and COLCIENCIAS (Colombia). 
J.R.P would like to thank the Theoretical Group at Berkeley
for their kind hospitality, as well as
the Jaime del Amo Foundation for a fellowship. Partial support
by US DOE under contract DE-AC03-76SF00098 is gratefully acknowledged.

\end{document}